\documentclass[review]{elsarticle}

\usepackage{lineno,hyperref}
\usepackage{subcaption}
\usepackage{amsmath,amssymb,amsfonts}
\usepackage{url}
\usepackage{wrapfig, framed, caption}
\usepackage[lined,ruled,linesnumbered]{algorithm2e}
\usepackage{algorithmic}
\usepackage{graphicx}
\usepackage{textcomp}
\usepackage{multirow}

\usepackage{booktabs}
\usepackage{array}
\usepackage{float}
\modulolinenumbers[5]

\journal{Journal of \LaTeX\ Templates}

\begin{document}

\begin{frontmatter}

\title{A Hybrid Parallelization Approach for Distributed and Scalable Deep Learning.}


\author[mymainaddress]{S. B. Akintoye}
\author[mymainaddress]{L. Han
\corref{mycorrespondingauthor}}
\cortext[mycorrespondingauthor]{Corresponding author}
\ead{l.han@mmu.ac.uk}

\author[mymainaddress]{ X. Zhang}
\author[mysecondaryaddress]{H. Chen}
\author[mysecondaryaddress2]{D. Zhang}


\address[mymainaddress]{Department of Computing and Mathematics, Manchester Metropolitan University, UK}
\address[mysecondaryaddress]{Department of Computer Science, University of Sheffield, UK}
\address[mysecondaryaddress2]{College of Computer Science and Technology, Nanjing University of Aeronautics and Astronautics, P.R.China}

\begin{abstract}
Recently, Deep Neural Networks (DNNs) have recorded significant success in handling medical and other complex classification tasks. However, as the sizes of a DNN model and the available dataset increase, the training process becomes more complex and computationally intensive, usually taking longer to complete. In this work, we have proposed a generic full end-to-end hybrid parallelization approach combining model and data parallelism for efficiently distributed and scalable training of DNN models. We have also proposed a Genetic Algorithm Based Heuristic Resources Allocation (GABRA) mechanism for optimal distribution of partitions on the available GPUs for computing performance optimization. We have applied our proposed approach to a real use case based on 3D Residual Attention Deep Neural Network (3D-ResAttNet) for efficient Alzheimer Disease (AD) diagnosis on multiple GPUs and compared it with the existing state-of-the-art  parallel methods. The experimental evaluation shows that our proposed approach outperforms the existing parallel methods and achieves almost linear speedup with little or no differences in accuracy performance when compared with the existing non-parallel DNN models.
\end{abstract}


\end{frontmatter}


\section{Introduction}\label{sec:introduction}
In recent time, Deep Neural Networks (DNNs) have gained popularity as an important tool for solving complex tasks ranging from image classification \cite{dl:Sen}, speech recognition \cite{dl:bbaschian}, medical diagnosis \cite{dl:Aysun, dl:Jojoa}, to the recommendation systems \cite{dl:Markus} and complex games \cite{Tembine, dl:Dennis}. However, training a DNN model requires a large volume of data, which is both data and computational intensive, leading to increased training time.\\
To overcome this challenge, various parallel and distributed computing methods \cite{pa:Jeffrey} have been proposed to scale up the DNN models to provide timely and efficient learning solutions. Broadly, it can be divided into data parallelism, model parallelism, pipeline parallelism and hybrid parallelism (a combination of data and model parallelism).

 Data parallelism is a parallelization method that trains replicas of a model on individual devices using different subsets of data, known as mini-batches \cite{pa:Krizhevsky}, \cite{dl:George}. In data parallel distributed training, each computing node or a worker contains a neural network model replica and a churn of dataset, and compute gradients which are shared with other workers and used by the parameter server to update the model parameters \cite{dp:Soojeong}. 
 Data parallelization is relatively easy to implement, and supported by all major frameworks such as Mxnet \cite{dl:Tianqi}, PyTorch \cite{PyTorch}, TensorFlow \cite{dl:Abadi}, and etc. However, as parameters increases,  the overhead for parameter synchronisation inceeases, leading to performance degradation.  In addition, when a DNN model size is too big, it couldn't be executed on a single device. Hence it is not possible to perform data parallelization. 

Model parallelism is a parallelization method where a large model is split, running concurrent operations across multiple devices with the same mini-batch \cite{pa:Jeffrey}. It can help to speed up the DNN training either through its implementation or algorithm. 
In model parallelism, each node or a worker has distinct parameters and computation of layer of a model, and also updates weight of allocated model layers. There is no need for synchronization of parameters values among the workers, often communication occurs when sharing  training outputs between  workers \cite{dl:Junya}. The major challenge of this technique is how to break the model into partitions which would be allocated to the workers \cite{dl:Ruben}. Moreover, using model parallelization alone does not scale well to a large number of devices \cite{pa:Azalia} as it involves heavy communication between workers. Pipelining parallelism splits the DNN models training tasks into a sequence of processing stages \cite{kim:pipe}. Each stage takes the result from the previous stage as input, with results being passed downstream immediately. However, one of the significant drawbacks of pipelining parallelism is that it is limited by the slowest stages and has limited scalability.

Recently, the combination of model and data parallelization methods known as Hybrid parallelization has been explored to leverage the benefits of both methods to minimize communication overhead in the multi-device parallel training of DNN models \cite{pa:Minjie}.
Song et al. \cite{dl:Song} proposed HyPar, a hybrid parallelism architecture for DNN training with an array of DNN accelerators, and has shown through experimental evaluation that HyPar outperformed either model parallelization or data parallelization alone. Furthermore, Ono et al. \cite{dl:Junya} proposed a hybrid data-model parallel approach for sequence-to-sequence (Seq2Seq) recurrent neural network (RNN) machine translation, where a model parallelization approach was applied to the RNN encoder-decoder part of the Seq2Seq model and a data-parallel approach was applied to the attention-softmax part of the model.\\
Despite the performance of the existing parallelization methods, they are still subject to further improvement by optimally allocating the model computations and data partitions to the available devices for better model training performance. In this paper, we have proposed a generic hybrid parallelization approach for parallel training of DNN in multiple Graphics Processing Units (GPUs) computing environments, which combines both model and data parallelization methods. Our major contributions are as follows:

\begin{itemize}
\item Development of a generic full end-to-end hybrid parallelization approach for the multi-GPU distributed training of a DNN model.

\item Model parallelization by splitting a DNN model into independent partitions, formulating the network partitions-to-GPUs allocation problem as a 0-1 multiple knapsack model, and proposing a Genetic Algorithm based heuristic resources allocation (GABRA) approach as an efficient solution to optimize the resources allocation.

\item Exploitation of data parallelization based on the All-reduced method and synchronous stochastic gradient descent across multiple GPUs for further acceleration of the overall training speed.

\item Evaluation of the proposed approach through a real use case study -- by parallel and distributed training of a 3D Residual Attention Deep Neural Network (3D-ResAttNet) for efficient Alzheimer's disease diagnosis.
\end{itemize}

The remainder of this paper is organized as follows: Section \ref{pa:Lit} reviews the related work of the study. Section \ref{pa:gpu} discusses the details of the proposed approach. In Section \ref{pa:exp}, the experimental evaluation is described. Section \ref{pa:con} concludes the work. 

\section{Related Work}\label{pa:Lit}

 This section provides an overview in relation to distributed training of deep neural networks and genetic algorithms for resource optimisation. 

\subsection{Parallel and Distributed Training of Deep Neural Networks (DNNs)}
As mentioned earlier, existing efforts on parallel and distributed training of DNNs can be broadly divided into three categories, which include data parallelism, model parallelism, pipeline parallelism and hybrid parallelism.

\subsubsection{Data Parallelism}
\begin{figure}
     \centering
     \begin{subfigure}[b]{0.49\textwidth}
         \includegraphics[width=\textwidth]{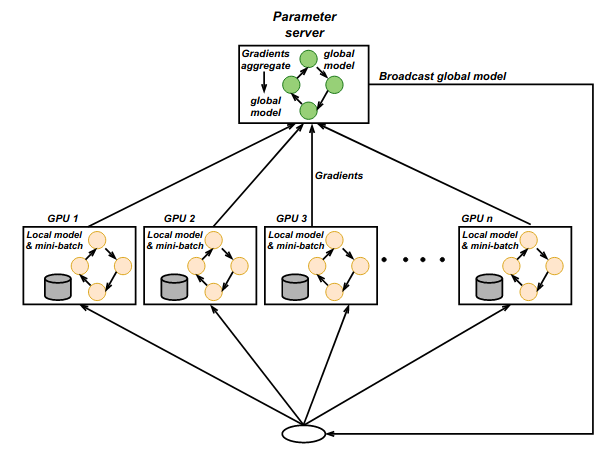}
         \caption{Parameter server architecture}
         \label{pd:par}
     \end{subfigure}
     \hfill
     \begin{subfigure}[b]{0.49\textwidth}
         \centering
         \includegraphics[width=\textwidth]{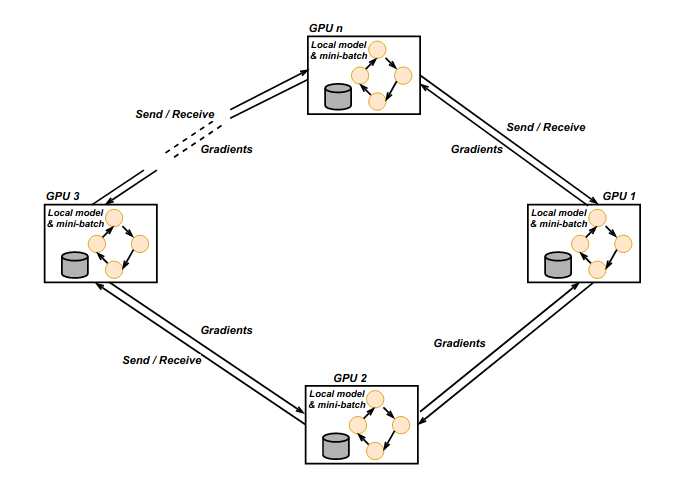}
         \caption{AllReduce architecture}
         \label{pd:rd}
     \end{subfigure}
             \caption{The conceptual data parallel training architectures}
        \label{fig:three graphs}
\end{figure}

In data parallelism,  a dataset is broken down into mini-batches and distributed across the multiple GPUs and each GPU contains a complete replica of the local model and computes the gradient. The gradients aggregation and updates among the GPUs are usually done either synchronously or asynchronously \cite{dp:Soojeong}. In synchronous training, all GPUs wait for each other to complete the gradient computation of their local models, then aggregate computed gradients before being used to update the global model. On the other hand, in asynchronous training, the gradient from one GPU is used to update the global model without waiting for other GPUs to finish. The asynchronous training method has higher throughput in that it eliminates the waiting time incurred in the synchronous training method. In both asynchronous and synchronous training, aggregated gradients can be shared between GPUs through the two basic data-parallel training architectures: parameter server architecture and AllReduce architecture. Parameter server architecture \cite{dl:Abadi} is a centralized architecture where all GPUs communicate to a dedicated GPU for gradients aggregation and updates. Figure \ref{pd:par} illustrates the parameter updates in data-parallel training, where each GPU computes the gradient from the local model and mini-batch. The parameter server aggregates gradients, and broadcasts the global model to all GPU for the next iteration processes. Alternately, AllReduce architecture \cite{dl:Sergeev} is a decentralized architecture where the GPUs share parameter updates in a ring network topology manner through the Allreduce operation as illustrated in Figure \ref{pd:rd}.

\subsubsection{Model Parallelism}
In model parallelization, model layers are divided into partitions and distributed across GPUs for parallel training \cite{dl:Zaharia, dl:Azalia}. In model parallel training, each GPU has distinct parameters and computation of the layer of a model, and also updates weight of allocated model layers. The intermediate data such as layer outputs and gradients for the forward pass and backward pass respectively are exchanged among GPUs. 
 Huo et al. \cite{dl:Zhouyuan} proposed a Decoupled Parallel Back-propagation (DDG), which splits the network into partitions and solves the problem of backward locking by storing delayed error gradient and intermediate activations at each partition. Similarly, Zhuang et al. \cite{dl:Huiping} adopted the delayed gradients method to propose a  fully decoupled training scheme (FDG). The work breaks a  neural network into several modules and trains them concurrently and asynchronously on multiple devices. 
However, the major challenges are how to break the model layers into partitions as well as the allocation of partitions to GPUs for efficient training performance \cite{dl:Ruben}. 

\subsubsection{Pipelining Parallelism}
Pipelining parallelism breaks the task (data and model) into a sequence of processing stages. Each stage takes the result from the previous stage as input, with results being passed downstream immediately. Various works have adopted this technique. Deep Stacking Networks (DSN) \cite{dl:Platt} overlapped neural network layer computations to optimise the performance of the training process. DSN consists of three modules, and each module computes a different hidden layer of the raw data. The raw data vector is concatenated with the output layer(s) in the lower module(s). The output of the lower module serves as input to the immediate higher module. Another deep learning system that uses the pipeline technique is PipeDream \cite{dl:Deepak}. Moreover, Lee et al. \cite{dl:Lee2} used the pipeline parallelism approach to overlap computation and communication for CNN training. They implement a thread in each computer server to spawn communication processes after the gradient is generated. Chen et al. \cite{dl:Chung3} proposed a pipelined model parallel execution method for high GPU utilisation and used a novel weight prediction technique to achieve a robust training accuracy.

\subsubsection{Hybrid Parallelism}
Several research works have explored both data and model parallelization methods for efficient DNN models training. Yadan et al. \cite{dl:Omry}  
achieved 2.2$\times$ speed-up when they trained a large deep convolutional neural network model with hybridized data and model parallelism.  Krizhevsky et al.\cite{pa:Krizhevsky} used model and data parallelization techniques to train a large deep convolutional neural network and classify 1.2 million high-resolution images in the ImageNet LSVRC-2010 contest into the 1000 different classes. Huang et al. \cite{pa:Yanping} proposed GPipe, a scalable model-parallelism library as well as a novel batch-splitting pipeline-parallelism algorithm to train large-scale convolutional and transformer-based neural networks. Shazeer et al. \cite{dl:Noam} proposed Mesh-TensorFlow where data parallelism is combined with model parallelism to improve training performance of transformer model with a huge number of parameters. In Mesh-TensorFlow, users split layers across the multi-dimensional mesh of processors and explored data parallelism technique in conjunction with the All-reduced update method. Moreover, Onoufriou et al. \cite{dl:Onoufriou} proposed Nemesyst, a novel end-to-end hybrid parallelism deep learning-based Framework, where model partitions are trained with independent data sets simultaneously. The application of the framework on the Recurrent Neural Network (RNN) models for retail refrigeration systems, achieved cogent performance and improved the scheduling of the defrosting cycles. Similarly, Oyama et al. \cite{dl:Oyama} proposed end-to-end hybrid-parallel training algorithms for large-scale 3D convolutional neural networks. The algorithms combine both data and model parallelisms to increase throughput and minimize I/O scaling bottlenecks. The results of performance evaluation with the CosmoFlow and 3D U-Net networks validate the potency of the algorithms. \\
The above-aforementioned approaches adopted data, model and pipeline parallelization separately or the combination of the methods to improve the performance of DNN models training. However, none of the existing methods considered the resource utilization and allocation problem in deep learning and provided solutions for efficient distributed training performance.

\subsection{Genetic Algorithms for Resource Management Optimization}
Resource management optimization is an important research topic in distributed computing systems \cite{pa:Zbigniew}. Several works have been proposed, with different techniques for addressing resource management problems, such as scheduling \cite{pa:Apolina} and allocation \cite{pa:Haji}. Genetic Algorithms (GA)s  are commonly used to optimize either homogenous or heterogeneous resources in distributed system environments \cite{pa:Zhang} \cite{Li-2018}. For instances, Gai et al. \cite{pa:Zhao} proposed the Cost-Aware Heterogeneous Cloud Memory Model (CAHCM) to provide high performance cloud-based heterogeneous memory service offerings. It proposed the Dynamic Data Allocation Advanced (2DA) algorithm based on genetic programming to determine the data allocations on the cloud-based memories for the model. 
Mezache et al. \cite{pa:Chaabane} proposed a resource allocation method based on GA to minimize the number of hosts required to execute a set of cloudlet associated with the corresponding set of the virtual machine, thereby reducing excessive power consumption in the data centre. Furthermore, Jiang et al. \cite{pa:Hui} developed a mathematical model to minimize the expected total makespan and the expected total cost of the disassembly service under the uncertainty nature of the disassembly process. They proposed a multi-objective genetic algorithm based on the non-dominated sorting genetic algorithm, which shows efficiency. Mosa and Sakellariou \cite{pa:Mosa} proposed a dynamic VM placement solution under consideration of heterogeneous resources. The solution used a GA to optimize the utilization of both CPU and memory with the aim to ensure better overall utilization in the cloud data centre. Devarasetty and Reddy \cite{pa:Devarasetty} proposed an optimization method for resource allocation in the cloud with the aim to minimize the deployment cost and improve the QoS performance. They used the GA to find optimal solutions to the allocation problem. 
In addition to resource allocation in the cloud environment, Mata and Guardieiro \cite{pa:Mata} investigated the resource allocation in the Long-Term Evolution (LTE) uplink and proposed a scheduling algorithm based on GA to find a solution for allocating LTE resource to the user requests. Moreover, Li and Zhu \cite{pa:Zhi} adopted genetic algorithm to develop a joint optimization method for offloading tasks to the mobile edge servers (MESs) in a mobile-edge computing environment under limited wireless transmission resources and MESs' processing resources. The outcomes of the simulation show that the proposed method can efficiently address the optimization problem of reducing the task offload completion time and resource allocation cost when compared with a single allocation offload strategy.

\section{The Proposed Approach}\label{pa:gpu}
In parallel and distributed computing, there are several considerations on the efficient training of DNN models including: 1) how to decompose a model or a dataset into parts/small chunks; 2) how to map and allocate these parts onto distributed resources for efficient computation as well as reducing communication overhead between computing nodes.  The number of network model partitions/data partitions/mini-batches will depend on the number of network layers and the number of available GPUs in a distributed computing environment. This involves partition strategy and resource allocation.
 
This work has proposed a generic full end-to-end hybrid parallelization approach for efficient training of a DNN model, which combines both data and model parallelization. 
For data parallelization, we have exploited data parallelization based on the All-reduced method and synchronous stochastic gradient descent across multiple GPUs for acceleration of the overall network training speed. For model parallelization, model layers are partitioned individually with the aim to reduce communication overhead during the training process. We have also designed a Genetic Algorithm-based heuristic resource allocation mechanism to map and allocate partitions to appropriate resources from efficient DNN training. 

\begin{figure*}[h!]
\centering
\includegraphics[scale=0.10]{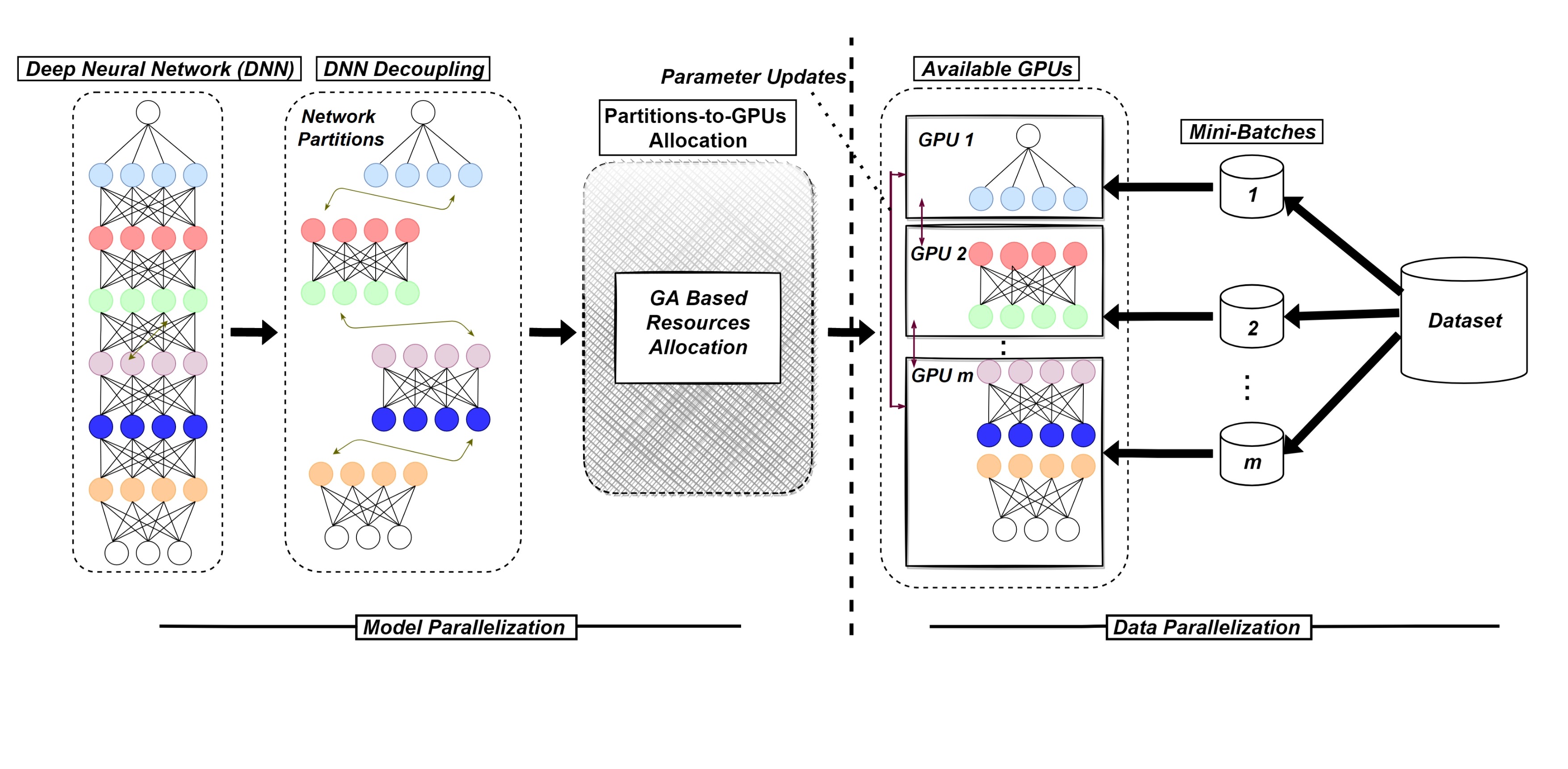} 
\caption{ The high level architecture of the proposed hybrid parallelization approach \label{pd:frame} }
\end{figure*}

Figure \ref{pd:frame} shows the high-level architecture, including 1) model parallelization consisting of network partitions and resource allocation components; and 2) data parallelization. The details of the proposed method are presented in the following sections. The important notations in this paper are detailed in Table \ref{tb:frame}.

\begin{table}
\begin{center}
\caption{Notations}
\label{tb:frame}
\begin{tabular}{{c}l*{3}{c}r{c}}
\toprule
Notations           & Descriptions   \\
\hline
$s_{i}$ & The $i-th$ Network layer. \\
$p_{n}$ &  The Network partitions of $n$ size.        \\
$Q$ & Number of Network layer.\\
$d_{m}$ &  The set of $m$ GPUs.\\
$t$ & Iteration.\\
$a^{t}$ & Activation at iteration $t$.\\
${w_{q}}$ & The weight parameter of $q$ layer.\\
$b$  & The Batch size. \\ 
$l$ & The Loss function.\\ 
$\gamma_{t}$ & Learning rate at iteration $t$.\\ 
$v$ & Data point.\\   
$V$ & Total data points of the dataset.\\  
$b$ & Batch size.\\       
$\Psi_{c}$ & The crossover operator. \\
$\Psi_{m}$ & The mutation operator. \\
$g^{t}_{s_{(i)}}$ & The gradient at partition $(i)$ at iteration $t$. \\ \bottomrule
\end{tabular}
\end{center}
\end{table} 

\subsection{Model Parallelization}
Model parallelization includes neural network model partitioning and Genetic Algorithm based heuristic resource allocation mechanism.
\subsubsection{Network Partitioning}
The principle of the network partitioning is based on the computation loads of each layer with the aim to reduce communication overhead during training process. The highly functional layers are partitioned individually as a single partition for even distribution of the DNN model layers. For instance, a convolution layer of CNN architecture has a large volume of weights and can be partitioned as a single partition for efficient parallel training performance. 

Specifically, let's assume a model network contains a set of layers $\lbrace s_{1}, \dots, s_{q}, \dots, s_{Q} \rbrace$. The model network $P$ is split into partitions $\lbrace p_{1}, p_{2},\dots,p_{n} \rbrace $ where $p_{i} = \lbrace s_{i}, s_{i} + 1,\dots,s_{i+1}-1 \rbrace$, $\lbrace s_{i}, s_{i} + 1,\dots,s_{i+1}-1 \rbrace$ denotes a set of layers in $i$ partition such that $1 \leq i \leq n$. In addition to this, all partitions are computed simultaneously, the gradient of the partition input is passed to the next partition $(i-1)$, while the partition output is sent to partition $(i+1)$ as its new input. In forward pass, the input $a^{t}_{s_{i-1}}$ from partition $(i-1)$ is sent to partition $i$ and gives activation $a^{t}_{s_{(i+1)}-1}$ at iteration $t$. Also, In backward pass, the $g^{t}_{s_{(i+1)}-1}$ denotes the gradient at partition $(i+1)$ at iteration $t$. For each layer $(s_{i} \leq q \leq s_{i+1} -1)$ such that $q \leq Q$, the gradient is given as: $\hat{g}^{t}_ {w_{q}}=  \dfrac{\delta a^{t}_{s_{(i+1)}-1}}{\delta w^{t}_{q} } g^{t}_{s_{(i+1)}-1}$ which can be updated by $w^{t+1}_{q} = w^{t}_{q} - \gamma_{t}\hat{g}^{t}_{w_{q}}$ where $\gamma_{t}$ is learning rate.

\begin{equation}
\hat{g}^{t-i+1}_ {w_{q}}=  \dfrac{\delta a^{t-i+1}_{s_{(i+1)}-1}}{\delta w^{t-i+1}_{q} } g^{t-i+1}_{s_{(i+1)}-1}
\end{equation}

which can be updated by:

\begin{equation}
w^{t-i+2}_{q} = w^{t-i+1}_{q} - \gamma_{t-i+1}\hat{g}^{t-i+1}_{w_{q}}
\end{equation}

where $\gamma_{t-i+1}$ is learning rate.\\

\subsubsection{Genetic Algorithm Based Resource Allocation (GABRA)}
To enable efficient DNN model training on multiple GPUs,  we have also proposed a Genetic Algorithm-based heuristic resource allocation mechanism.  We formulate the problem of allocating GPUs to network partitions as a 0-1 multiple knapsack problem model. As previously illustrated, we consider computation load of a set of partitions $p_{i}$, where $i = \lbrace 1, 2, \dots,n \rbrace$. We also consider the capacity of a set of available GPUs $G$, each denoted by $d_{j}$ where $j = \lbrace 1, 2, \dots,m \rbrace$ and $d_{j} \in G$. Furthermore, we assume that each GPU runs at least one partition and each partition need to be allocated to only one GPU.\\
Let $C = (c_{ij}) \in \mathbb{R}^{n \times m}$  be a $n \times m$ matrix in which $c_{ij}$ is a profit of allocating GPU $i$ to partition $j$:

\begin{equation}
 c_{ij} = \dfrac{p_{i}}{d_{j}}
\end{equation}

Also, Let $X = (x_{ij}) \in \mathbb{R}^{n \times m}$ where

\begin{equation}
x_{ij} = \begin{cases}
1,& $if gpu i is allocated to partition j$ \\
0,& $otherwise$ .
\end{cases}
\end{equation}

Thus, we formulate the multiple knapsack model in terms of a function $z$ as:

\begin{equation}
\label{eq7}
 \max_{p, x,c}\  z(X) = \sum\limits_{i=1}^{n}\sum\limits_{j=1}^{m}x_{ij}c_{ij} 
\end{equation}

subject to:

\begin{equation}
\label{eq8}
   \begin{array}{rcl} 
      
      \sum\limits_{i=1}^{n}p_{i}x_{ij} &\leq& d_{j},     \forall j \in M =  \lbrace 1, 2,\ldots m \rbrace \\
      \end{array}
   \end{equation}

\begin{equation}
\label{eq9}
 \begin{array}{rcl} 
\sum\limits_{j=1}^{m}x_{ij} &=& 1,     \forall  i  \in N = \lbrace 1, 2,\ldots n \rbrace \\
\end{array}
\end{equation}

\begin{align}
\label{eq10}
\begin{array}{rcl} x_{ij} &=& \mbox{0 or 1},   \mbox{  for i = 1, 2,\ldots n}, \mbox{ and for j = 1, 2,\ldots m} \end{array}
\end{align}
Our goal is to find ( \ref{eq10}) that guarantees no GPU is overutilized and yields the maximum profit simultaneously. Thus, the objective function in equation \ref{eq7} maximizes the sum of the profits of the selected partitions. The constraint Eq. \ref{eq8} implies that each partition is allocated to at most one GPU, while constraints Eq. \ref{eq9} ensures that the capacity of each available GPU is not exceeded.\\
Next, we present a Genetic Algorithm-Based Resources Allocation (GABRA) as an efficient solution to the model. Genetic Algorithm has been proven as a stochastic method to produce high-quality solutions for solving combinatorial optimization problems, particularly NP-hard problems \cite{pa:Todd}, \cite{pa:Bader}.
\begin{algorithm}
\SetKwInOut{Input}{input}\SetKwInOut{Output}{output}
\SetKw{KwGoTo}{go to}
\Input{  $\lbrace p_{1}, p_{2},....,p_{n} \rbrace$ : computation loads of partitions \\
     $\lbrace d_{1}, d_{2},....,d_{m} \rbrace$ : capacity values of available GPUs} 
\Output{optimized solution $f(Z^{*})$} 
evaluate $c_{ij} = \dfrac{p_{i}}{d_{j}}$, for $i = \lbrace 1, 2,\ldots,n \rbrace$ and  $j = \lbrace 1, 2,\ldots,m \rbrace$\; 
set $t = 0$\;
initialise $\mathcal{P}(t) = \lbrace  \beta_{1}, \beta_{2},\ldots, \beta_{n}  \rbrace$ \label{inia}\;
evaluate $\mathcal{P}(t):\lbrace  f(\beta_{1}), f(\beta_{2}),\ldots, f(\beta_{n})  \rbrace$ \;
find $Z^{*} \in  \mathcal{P}(t)$ such that $f(Z^{*}) \geq f(Z), \forall Z \in {P}(t)$  \;
\While {$(t < t_{max})$}{ 
select $\lbrace Y_{1}, Y_{2} \rbrace$ = $\phi(\mathcal{P}(t))$;\ //  $\phi$ is a selection function \label{marker}\;
crossover $W = \Psi_{c}(Y_{1}, Y_{2})$; //$\Psi_{c}$ is a crossover function\;
mutate $W \longleftarrow \Psi_{m}(W)$; //$\Psi_{m}$ is a mutation function\;
\If { $W$ = any $Z \in \mathcal{P}(t)$}
{ingore $W$ and \KwGoTo \ref{marker}}
evaluate $f(W)$  \;
find $Z^{'} \in \mathcal{P}(t)$ such that $f(Z^{'}) \leq f(Z), \forall Z \in {P}(t)$ and replace $Z^{'} \longleftarrow W$  \label{rep}\;
\If {$f(W) > f(Z^{*})$} {$Z^{*} \longleftarrow W$; //update best fit $Z^{*}$} 
$ t \longleftarrow t + 1$ \label{marker2}\;

}
\Return $Z^{*}$, $f(Z^{*})$
 \caption{Genetic Algorithm Based Resources Allocation (GABRA) \label{algo:rand}}
\end{algorithm}
The Algorithm \ref{algo:rand} shows the pseudo-code of the GABRA for solving GUPs-to-partitions allocation problem. It consists of four major parts: \textit{input}, \textit{initialization}, \textit{looping} and \textit{output}. In the initialization part (line \ref{inia}), unlike the classical GA, the set of chromosomes which also known as initial population $\mathcal{P}(t)$ for allocating GPUs to partitions, is generated as indicated in the Algorithm \ref{algo:rand}, by randomizing the allocation of resources without exceeding their capacities with respect to the computation load of each network partition. \\
\begin{algorithm}
\SetKwInOut{Input}{input}\SetKwInOut{Output}{output}
\Input{  $\lbrace p_{1}, p_{2},....,p_{n} \rbrace$ : computation loads of partitions \\
     $\lbrace d_{1}, d_{2},....,d_{m} \rbrace$ : capacity values of available GPUs} 
\Output{initial population} 
\For{(all partition loads)}{ 
 randomize the allocation of partitions to the number of the available GPUs }
 \Return initial population 
 \caption{initial population algorithm \label{algo:rand}}
\end{algorithm}  
The looping part contains fitness evaluation, selection, crossover and mutation functions. The objective is to optimize the total profit of allocating GPUs to partitions. The fitness evaluation validates the optimal solution condition with respect to the optimization objectives. Thus, the fitness value of each chromosome is calculated as:

\begin{equation}
f(\beta) = \sum\limits_{i=1}^{n}c_{ij}\beta_{i}, \mbox{ and for j = 1, 2,\ldots, m}
\end{equation}

In the case where the optimal solution condition is not satisfying the optimization objectives, a new population is computed from an initial population of the solutions using their fitness values and genetic functions: selection, crossover and mutation functions in the looping part (lines \ref{marker} - \ref{marker2}).
We use the selection function ($\phi$), which is based on the roulette wheel method \cite{Yu:Dong} to select the best chromosomes. The selection is based on the chromosomes' fitness values, representing the total profit of allocating partitions to the available GPUs. The chromosomes with higher fitness values are selected for the generation of the next population. The midpoint crossover function $\Psi_{c}$ as described in Algorithm \ref{algo:cross}, works on two-parent chromosomes $\lbrace Y_{1}, Y_{2} \rbrace$ with crossover probability 0.8 and produces a new individual.\\

\begin{algorithm}
\SetKwInOut{Input}{input}\SetKwInOut{Output}{output}
\Input{$Y_{1}, Y_{2}$ : two parent chromosomes }
\Output{$Y_{\lambda1}$, $Y_{\lambda2}$ : two offspring chromosomes} 
$\Phi$ = $length(Y_{1})$\;
$cp$ = $\frac{Y_{1}}{2}$; //mid cross point\;
$Y_{\lambda1}$ = $Y_{1}(1:cp) \cup Y_{2}(cp:\Phi)$\;
$Y_{\lambda2}$ = $Y_{1}(cp:\Phi) \cup Y_{2}(1:cp)$\;
\Return $Y_{\lambda1}$, $Y_{\lambda2}$
\caption{Crossover function ($\Psi_{c}$) \label{algo:cross}}
\end{algorithm}  

Next, the inversion mutation functions $\Psi_{m}$ is adopted where a subset of genes in a chromosome is selected and inverted to form mutated offspring. In the line \ref{rep}, the old chromosomes in the current population are replaced with the new chromosomes to form a new population. Finally, the algorithm terminates when the maximum number of generations is reached, or the optimal total profit of allocating GPUs-to-partitions is obtained.

\subsection{Data Parallelization}
\begin{figure}[h!]
\centering
\includegraphics[scale=0.27]{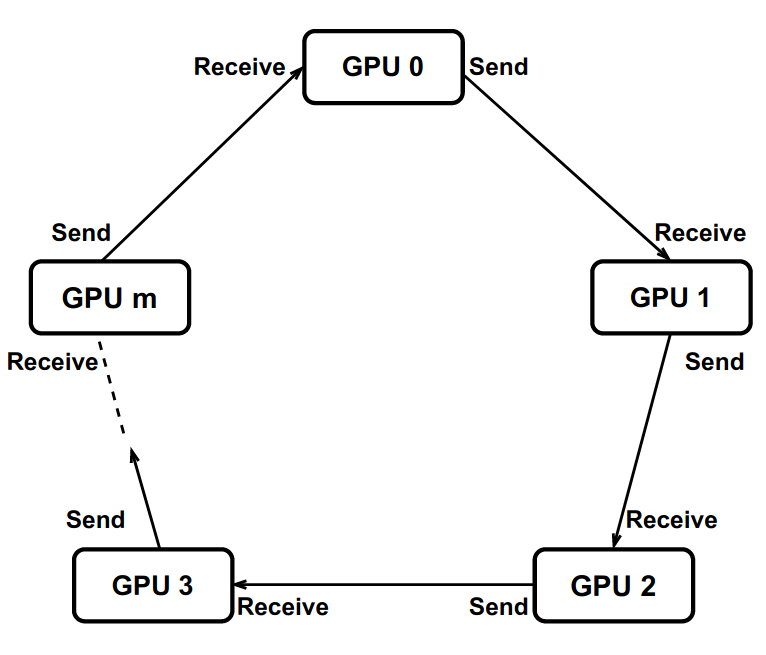} 
\caption{ Ring All-reduce topology \label{pd:top} }
\end{figure}

To accelerate each training step, we adopt Asynchronous Stochastic Gradient Descent (ASGD) \cite{pa:Jeffrey} as well as ring All-reduce mechanisms for parameters updates \cite{dl:Yuan}. ASGD achieves a faster training speed as there is no need of waiting for the slowest GPU in every iteration for the global model updates. The ring All-reduce is an optimal communication algorithm to minimize the communication overhead among the GPUs \cite{Pitch}. As shown in Fig. \ref{pd:top}, all GPUs are logically arranged in a ring All-reduce topology, Each GPU sends and receives required information to update its model parameters, from the neighbour GPUs.  

In all, the objective is to minimize as follows:

\begin{equation}
f(w;V) = \dfrac{1}{b \times m}\sum\limits_{i=1}^{b \times m}\ell(w,v_{i})
\end{equation}

where $f$ is a neural network, $b$ is the batch size, $m$ is the number of GPUs, $\ell$ is a loss function for each data point $v \in V$, and $w$ is the trainable parameter of the neural network.\\
The derivative of this objective which also referred to as the gradient is given as: 

\begin{equation}
\frac{\partial f(w;V)}{\partial w} = \dfrac{1}{b \times m}\sum\limits_{i=1}^{b \times m} \frac{\partial \ell(w,v_{i})}{\partial w}
\end{equation}

In data parallellization, the gradient updates is calculated as a sum of summations, each of which is the sum of derivatives over $b$ data points, and is given as:

\begin{equation}
\begin{aligned}
\frac{\partial f(w;V)}{\partial w}=\dfrac{1}{m}\left( \dfrac{1}{b}\sum\limits_{i=1}^{b} \frac{\partial \ell(w,v_{i})}{\partial w} +  \dfrac{1}{b}\sum\limits_{i=b+1}^{b \times 2} \frac{\partial \ell(w,v_{i})}{\partial w} +     \right) \\  \left(\ldots +  \dfrac{1}{b}\sum\limits_{i=b \times (m-1)+1}^{b \times m} \frac{\partial \ell(w,v_{i})}{\partial w} \right)
\end{aligned}
\end{equation}

In addition, the speed of data-parallel training with $m$ GPUs can be expressed as: 
\begin{equation}
ST_{m} = \frac{T_{1}}{T_{m}} \times \frac{TS_{1}}{TS_{m}} \times \frac{E_{1}}{E_{m}}
\end{equation}

where $T_{1}$ is the average training time per step for using one GPU, while $T_{m}$ is the time per step for using $m$ GPUs. $E_{1}$ is the number of epochs required to converge for one GPU, while $E_{m}$ is the number of epochs required for $m$ GPUs.

\section{Experimental Evaluation Through a Real Use Case Study}\label{pa:exp}
We have applied our approach to a real case study in Neurocomputing to evaluate the effectiveness of the proposed method in this work.  Previously, we have developed a 3D explainable residual self-attention convolutional neural network (3D-ResAttNet) to automatically classify discriminative atrophy localization on sMRI image for Alzheimer's Disease (AD) diagnosis \cite{Xingye}. It is a non-parallel model and runs on a single GPU. To evaluate the proposed parallel approach, we have parallellized our previous 3D-ResAttNet model and ran it on a multiple-GPUs setting comparing the performance with and without parallelization.  

Moreover, we have compared our approaches with the state-of-the-art methods including Distributed Data Parallel (DDP) and Data Parallel (DP) from PyTorch framework \cite{pa:Shen}, FDG \cite{dl:Zhouyuan} and DDG \cite{dl:Huiping}.


\subsection{Evaluation Metrics}
We have adopted standard metrics for performance evaluation including Speedup (S), Accuracy (ACC) and Training Time (TT). The Speedup (S) is to measure the scalability and computing performance. It is defined as the ratio of the serial runtime of the best sequential algorithm for solving a problem to the time taken by the parallel algorithm to solve the same problem on multiple processors (e.g., GPUs in this case). It can be calculated as: 

\begin{equation}
S = T_{s}/T_{p}
\end{equation}

where $T_{s}$ represents computing time on a single machine or GPU. $T_{p}$ refers to the computing time on multiple machines or GPUs. The Accuracy (ACC) measures the classification accuracy and is
defined as:

\begin{equation}
ACC = (TP + TN)/(TP + TN + FP + FN) 
\end{equation}

where TP = True positive, FP = False positive, TN = True negative and FN = False negative. Training Time (TT) is the time taken for training 3D-ResAttNet using the proposed approach and other existing distributed training methods.

\subsection{System Configuration}
We have conducted our experiments on an Amazon Web Service (AWS) EC2 P3 instance. Specifically, we used a \textit{p3.16xlarge} instance consisting of 8 NVIDIA Tesla V100 GPUs developed purposely for the deep learning and Artificial intelligent crowd to provide ultra-fast GPU to GPU communication through NVLink technology. Other hardware configuration of the \textit{p3.16xlarge} instance includes 128GB GPU memory, 64 vCPUs, 488GM memory, and 25Gbps network bandwidth. Additionally, software configuration /installation include: Ubuntu 18.04, Python 3.7.3, Pytorch 1.2.0, Torchvision 0.4.0, Numpy 1.15.4, Tensorboardx 1.4, Matplotlib 3.0.1, Tqdm 4.39.0, nibabel, fastai, and NVIDIA Collective Communications Library (NCCL) CUDA toolkit 10.2 - a library of multi-GPU collective communication primitives \cite{Nvidia}.

 \begin{figure}[h!]
\centering
\includegraphics[scale=0.33]{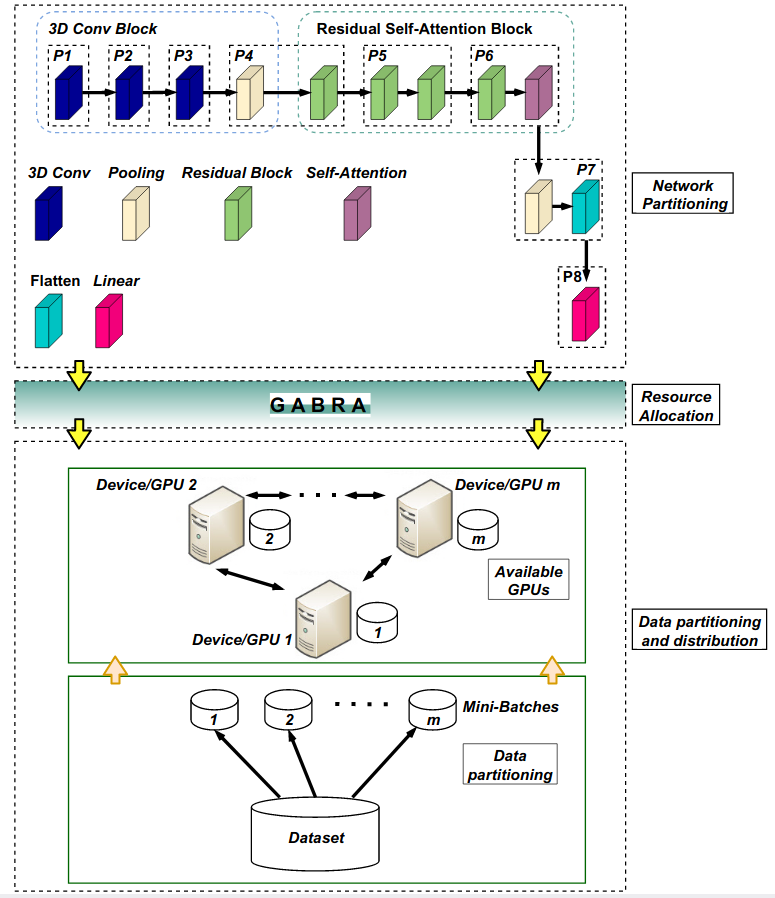} 
\caption{ The hybrid parallelisation of 3D-ResAttNet \label{pd:dl} }
\end{figure}

\subsection{A Use Case - Parallelization of 3D-ResAttNet For Alzheimer's Disease (AD) Diagnosis}
As described earlier, we have applied our hybrid parallelization approach to our previous non-parallel 3D-ResAttNet for automatic detection of the progression of AD and its Mild Cognitive Impairments (MCIs) such as Normal cohort (NC), Progressive MCL (pMCI) and Stable MCI (sMCI) from sMRI scans \cite{Xingye}.  It includes two types of classification: NC vs. AD, and pMCI vs. sMCI. 

\subsubsection{The High-Level Parallelization Of The System}

 Fig. \ref{pd:dl} shows the high level parallelization of our previous 3D-ResAttNet model architecture based on self-attention residual mechanism and explainable gradient-based localisation class activation mapping (Grad-CAM) to improve AD diagnosis performance. The 3D-ResAttNet model consists of 3D Convolutional blocks (Conv blocks), Residual self-attention block, and Explainable blocks. Conv blocks use a 3D filter for computation of the low-level feature representations. The residual self-attention block combines two important network layers: Residual network layer and Self-attention layer. The residual network layer comprises  two Conv blocks consisting of $3 \times 3 \times 3$ 3D convolution layers, 3D batch normalization and rectified-linear-unit nonlinearity layer (ReLU). The explainable block uses 3D Grad-CAM to improve the model decision.  
 
 As shown in  Fig. \ref{pd:dl}, the hybrid parallelization approach for 3D-ResAttNet is divided into three phases: the splitting of 3D-ResAttNet into partitions, allocation of GPUs to partitions, and data partitioning and distribution. For data parallelization, we adopt asynchronous stochastic gradient descent as well as ring All-reduce mechanisms for parameters updates and equally distribute data parts to each GPU.  For model parallelization, the network model is partitioned based on the computational complexity which usually synonymous to the number of basic operations, such as multiplications and summations, that each layer performs. Each Conv Block in the network consists of a $3 \times 3 \times 3$ 3D convolution layer, 3D batch normalization and a rectified-linear-unit nonlinearity layer (ReLU). Moreover, a convolutional layer has higher operations with complexity $\mathcal{O}(C_{o}.C_{1}.T.H.W.K_{T}.K_{H}.K_{W})$ where $C_{o}$ and $C_{1}$ denote the number of output and input channels respectively, $T$, $H$ and $W$ are image dimension, and $K_{T}$, $K_{H}$ and $K_{W}$ are filter dimension. Consequently, we partitioned each Conv block individually as a single partition while other layers with less computation operations are partitioned as shown in Fig. \ref{pd:dl}.\\

\begin{table}[h!]
\caption{Demographic data for the subjects from ADNI database}
\label{tab3:table0}
\begin{tabular}{p{0.9cm}>{\centering\arraybackslash}p{1.9cm}>{\centering\arraybackslash}p{1.9cm}>{\centering\arraybackslash}p{1.9cm}>{\centering\arraybackslash}p{1.8cm}}
\toprule
Class            & Number/Size & Gender (Male/Female) & Age (Mean/Std) & MMSE (Mean/Std) \\
\hline
AD & 389/1.4GB & 202/187 & 75.95/7.53 & 23.28/2.03    \\ 
pMCI  & 172/484MB          & 105/67 & 75.57/7.13 & 26.59/1.71    \\ 
sMCI & 232/649MB          & 155/77 & 75.71/7.87 & 27.27/1.78  \\ 
NC &400/2.4GB   & 202/198 & 76.02/5.18 & 29.10/1.01   \\ \bottomrule
\end{tabular}
\end{table}

\subsubsection{Dataset}
The dataset is obtained from the Alzheimer's Disease Neuroimaging Initiative (ADNI) database (\url{http://adni.loni.usc.edu}), which is the same dataset previously used for validation of our 3D-ResAttNet. The ADNI was developed in 2003 by Dr Michael W.Weiner under the public-private partnership with the aim to test the veracity of combining serial magnetic resonance imaging (MRI), positron emission tomography (PET), other biological markers, and clinical and neuropsychological assessment in measuring the progression of mild cognitive impairment (MCI) and early Alzheimer's disease (AD). The dataset contains 1193 MRI scans of four classes: 389 Alzheimer's Disease (AD), 400 Normal Cohort (NC), 232 sMCI and 172 pMCI patients. The demographic data for this dataset is shown in Table \ref{tab3:table0}.

\begin{table*}[h!]
\begin{center}
\caption{Parallel training performance of 3D-ResAttNet using our proposed approach} 
\label{tab3:table1}
\begin{tabular}{|>{\centering}p{1.1cm}|>{\centering}p{0.8cm}|>{\centering}p{0.8cm}|>{\centering}p{0.8cm}|>{\centering}p{0.8cm}|>{\centering}p{0.8cm}|>{\centering}p{0.8cm}|>{\centering}p{0.8cm}|>{\centering}p{0.8cm}|>{\centering}p{0.8cm}|>{\centering}p{0.8cm}|>{\centering}p{0.8cm}|>{\centering}p{0.8cm}|>{\centering}p{0.8cm}|>{\centering}p{0.8cm}|} \hline
&\multicolumn{4}{ c| }{3D-ResAttNet18} &  \multicolumn{4}{ c| }{3D-ResAttNet34} \\      \cline{2-9}
 &\multicolumn{2}{ c| }{sMCI vs. pMCI} &  \multicolumn{2}{ c| }{AD vs. NC}&  \multicolumn{2}{ c| }{sMCI vs. pMCI}&  \multicolumn{2}{ c| }{AD vs. NC} \\ \cline{2-9}
\#GPUs &ACC&TT (mins)& ACC&TT (mins)& ACC&TT (mins)& ACC&TT (mins)
 \tabularnewline \hline
 1 & 0.82 & 34 & 0.95 & 62 & 0.83 & 37 & 0.96 & 68\tabularnewline \hline
 2 & 0.81 & 15 & 0.94 & 26 & 0.84 & 17 & 0.96 & 29\tabularnewline \hline
 3 & 0.81 & 13 & 0.96 & 21 & 0.84 & 14 & 0.97 & 24\tabularnewline \hline
 4 & 0.82 & 11 & 0.92 & 18 & 0.83 & 12 & 0.96 & 21\tabularnewline \hline
 5 & 0.8 & 10 & 0.94 & 17 & 0.84 & 11 & 0.97 & 19\tabularnewline \hline
 6 & 0.82 & 9 & 0.95 & 15 & 0.82 & 10 & 0.96 & 16\tabularnewline \hline
 7 & 0.81 & 7 & 0.95 & 12 & 0.84 & 8 & 0.96 & 14\tabularnewline \hline
 8 & 0.81 & 6 & 0.95 & 11 & 0.84 & 7 & 0.96 & 12 \tabularnewline \hline
\end{tabular}
\end{center}
\end{table*}
\begin{figure*}[h!]
     \begin{subfigure}[b]{0.49\textwidth}
         \includegraphics[width=\textwidth]{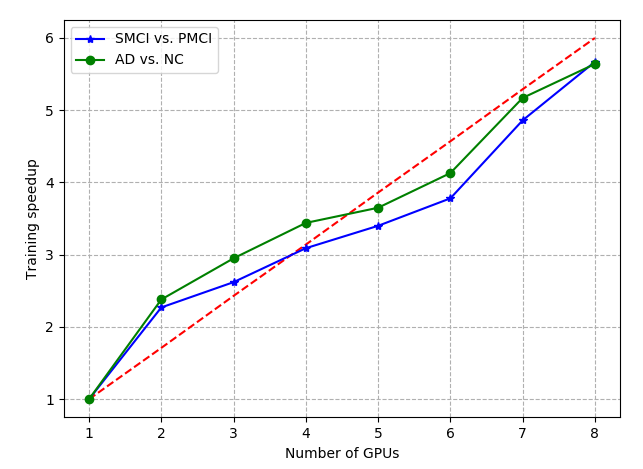}
         \caption{sMCI vs. pMCI and AD vs. NC classification with respect to 3D-ResAttNet18}
         \label{fig2_1}
     \end{subfigure}
     \hfill
     \begin{subfigure}[b]{0.49\textwidth}
         \includegraphics[width=\textwidth]{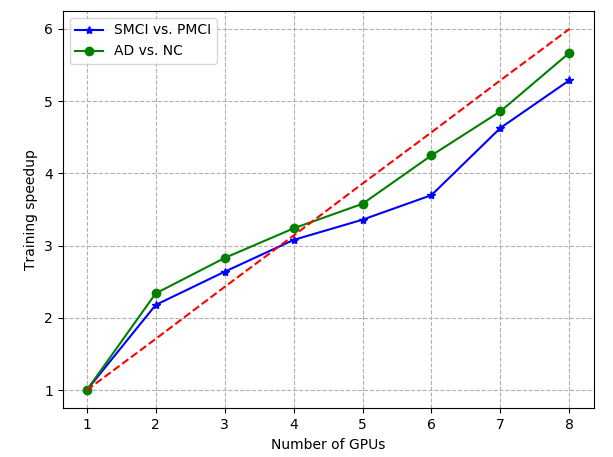}
         \caption{sMCI vs. pMCI and AD vs. NC classification with respect to 3D-ResAttNet34}
         \label{fig2_2}
     \end{subfigure}
    \hfill
     \begin{subfigure}[b]{0.49\textwidth}
         \includegraphics[width=\textwidth]{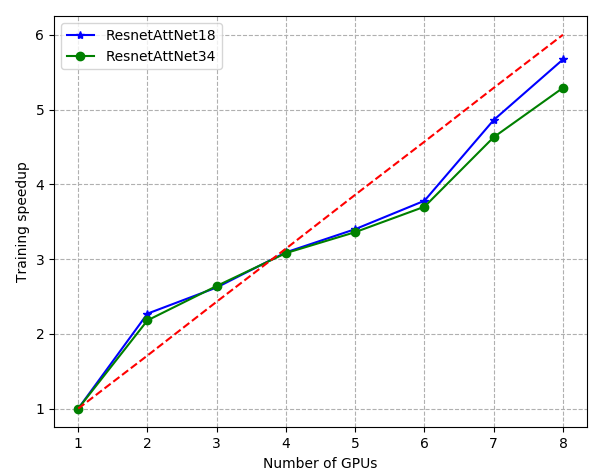}
         \caption{3D-ResAttNet18 vs. 3D-ResAttNet34 with respect to sMCI vs. pMCI}
         \label{fig2_3}
     \end{subfigure}
     \hfill
     \begin{subfigure}[b]{0.49\textwidth}
         \includegraphics[width=\textwidth]{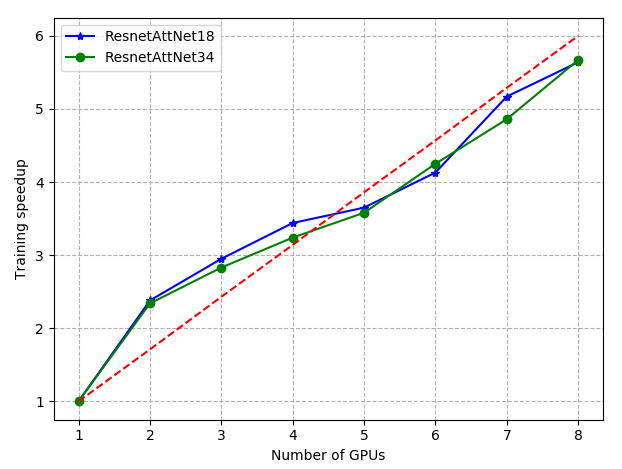}
         \caption{3D-ResAttNet18 vs. 3D-ResAttNet34 with respect to AD vs. NC}
         \label{fig2_4}
     \end{subfigure}
\caption{Speedup of our proposed approach \label{fig2}}
        
\end{figure*}

\begin{figure*}[h!]
     \begin{subfigure}[b]{0.49\textwidth}
         \includegraphics[width=\textwidth]{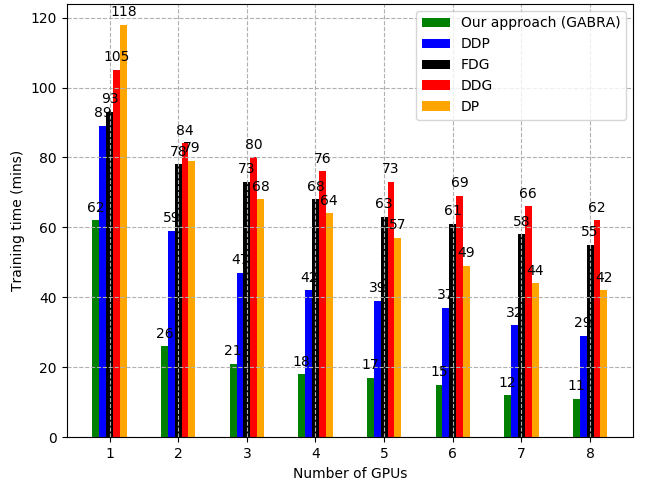}
         \caption{Training Time of 3D-ResAttNet18 over AD vs. NC Dataset}
         \label{fig2_5}
     \end{subfigure}
     \hfill
     \begin{subfigure}[b]{0.49\textwidth}
         \includegraphics[width=\textwidth]{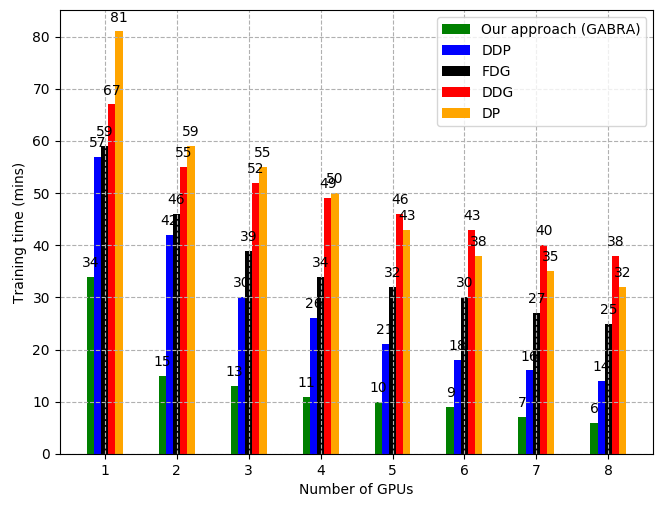}
         \caption{Training Time of 3D-ResAttNet18 over sMCI vs. pMCI Dataset}
         \label{fig2_6}
     \end{subfigure}
    \hfill
     \begin{subfigure}[b]{0.48\textwidth}
         \includegraphics[width=\textwidth]{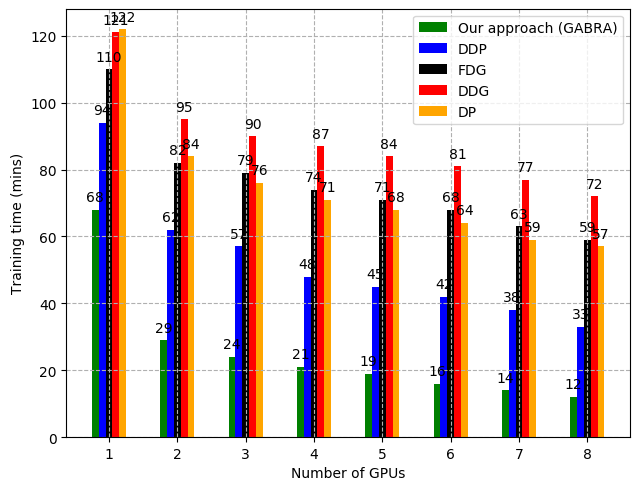}
         \caption{Training Time of 3D-ResAttNet34 over AD vs. NC Dataset}
         \label{fig2_7}
     \end{subfigure}
     \hfill
     \begin{subfigure}[b]{0.49\textwidth}
         \includegraphics[width=\textwidth]{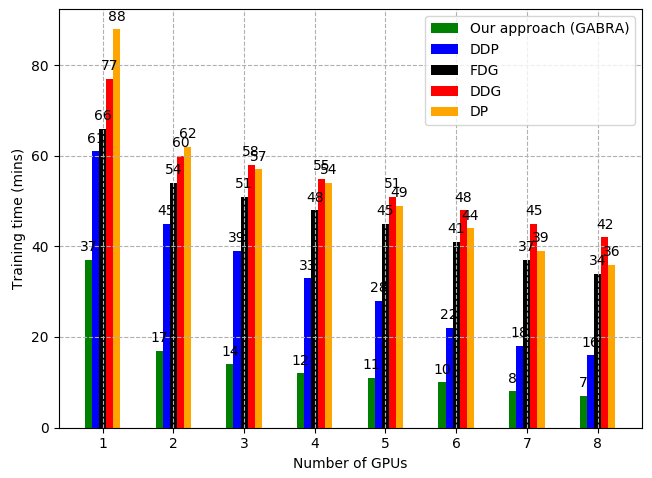}
         \caption{Training Time of 3D-ResAttNet34 over sMCI vs. pMCI Dataset}
         \label{fig2_8}
     \end{subfigure}
\caption{Training time of proposed approach, FDG, DDG, DDP and DP \label{fig2}}
        
\end{figure*}

\subsection{Experiments}
We have conducted experiments under different strategies:
\begin{enumerate} [1)]

\item We have evaluated the model performance in the parallel setting across the number of GPUs.   
\item We have compared our proposed approach with four existing parallel methods including: two PyTorch generic distributed training methods: $DistributedDataParallel$ (DDP) and $DataParallel$ (DP), and delayed gradient parallel methods: FDG and DDG  for further evaluation. DDP is a multi-process data parallel training across GPUs either on a single machine or multiple machines, while DP is for single-process multi-parallel training using multiple GPUs on a single machine \cite{pa:Shen}. Both FDG and DDG were implemented by partitioning data and trained parallel models with sub-data across multiple GPUs.    
\end{enumerate}

In all experiments, we carried out several distributed training of 3D-ResAttNet18 and 3D-ResAttNet34 for two classification tasks: sMCI vs. pMCI and AD vs. NC with different number of GPUs (ranging from 1 to 8). Furthermore, we optimized model parameters with SGD, a stochastic optimization algorithm, and adopted other training parameters including a batch size of six samples, cross-entropy as the loss function, and 50 epochs for better convergence. In addition, we set the initial learning rate (LR) as $1 \times 10^{-4}$, then reduced it by $1 \times 10^{-2}$ with increased iterations.


\subsection{Experimental Results and Discussion}
\subsubsection{Performance of 3D-ResAttNet in the parallel setting }

We conducted the experiments on 3D-ResAttNet model for two classification tasks: sMCI vs. pMCI and AD vs. NC.  Table \ref{tab3:table1} shows the experiment results of our parallel 3D-ResAttNet (with 18 and 34 layers respectively) in terms of both training time (TT) and accuracy.  Based on it, we calculated the speedup,  as shown in Figs. \ref{fig2_1}, \ref{fig2_2}, \ref{fig2_3}, and \ref{fig2_4}. 

 In all cases, it is observed that our proposed approach achieves almost linear speedup, which demonstrates the scalability of our approach in that the number of GPUs is directly proportional to the training speedup performance. For instance, in AD vs. NC classification task with 3D-ResAttNet34, the training speedup for 1, 2 3, 4, 5, 6, 7, and 8 GPUs are 1, 2.38, 2.95, 3.44, 3.65, 4.13, 5.17, and 5.64 respectively. A similar trend is also observed in the sMCI vs. pMCI classification task with 3D-ResAttNet34, the training speedup for 1, 2 ,3, 4, 5, 6, 7, and 8 GPUs are 1, 2.27, 2.62, 3.09, 3.40, 3.78, 4.86, and 5.67 respectively.

\subsubsection{Comparison Study Between Our Proposed Approach And The Existing Parallel Approaches (DDP, DP, DDG and FDG)}

 We have compared our approach with DDP, DP, DD and FDG, and the experiment results are shown in Figs. \ref{fig2_5}, \ref{fig2_6}, \ref{fig2_7}, and \ref{fig2_8}.  It can be seen that our proposed approach outperforms the existing methods in terms of training time. For instance, for 3D-ResAttNet18 on AD vs. NC and sMCI vs. pMCI classification tasks, the training time incurred by our proposed approach is averagely more than 80\% lower than DDP, DP, DDG and FDG. Similarly, there are related trends when comparing the proposed approach with the DDP, DP, DDG and FDG with respect to the distributed training of 3D-ResAttNet34 for two classification tasks: AD vs. NC and sMCI vs. pMCI

\subsubsection{Accuracy Comparison With The Existing Works}
Table \ref{tab6:table1} shows the accuracy comparison results from seven state-of-the-art deep neural networks and our methods. The best testing accuracies obtained in our approach are 0.97\% and 0.84\% for AD vs. NC and sMCI vs.pMCI classification respectively. The results show that our proposed approach performs efficiently when compared with the existing works in terms of accuracy. In addition, our work implements parallel distributed training of networks in a multi-GPU environment, whereas existing works are non-parallel methods.
 \begin{table}[h!]
\begin{center}
\caption{Accuracy comparison with the existing works}
\label{tab6:table1}
\begin{tabular}{p{3.3cm}>{\centering}p{2.7cm}>{\centering}p{1.5cm}>{\centering}p{1.4cm}>{\centering\arraybackslash}p{1.3cm}}
\toprule
References & Model & Parallel Training&sMCI vs. pMCI & AD vs. NC \\
\hline
Hosseini-Asl et al \cite{pa:Robert} & CNN &No& N/A & 0.97    \\ 
Suk et al.\cite{pa:Suk}  &DBM          & No&0.76 & 0.95 \\ 
Sarraf et al.\cite{pa:Sarraf} & CNN          &No& N/A & 0.96  \\ 
Billones et al.\cite{pa:Billones} &CNN   &No& N/A & 0.91    \\ 
Li et al.\cite{pa:Hongming} &3D CNN    &No& 0.77 & 0.91   \\ 
Shi et al.\cite{pa:shi} &MM-SDPN   &No& 0.75 & 0.95\\
Zhang et al.\cite{Xingye} &3D-ResAttNet34   &No& 0.84 & 0.97    \\ 
Our approach &3D-ResAttNet34&Yes&0.84&0.97 \\
 \bottomrule
\end{tabular}
\end{center}
\end{table}

\section{Conclusion and Future Work}\label{pa:con}
In this work, we have proposed a hybrid parallelization approach that combines both model and data parallelization for parallel training of a DNN model. The Genetic Algorithm based heuristic resources allocation mechanism (GABRA) has also been developed for optimal distribution of network partitions on the available GPUs for performance optimization. Our proposed approach has been compared with the existing state-of-the-art parallel methods and evaluated with a real use case based on our previous 3D-ResAttNet model developed for efficient AD diagnosis. The experiment results show that the proposed approach achieves linear speedup, which demonstrates its scalability and efficient computing capability with little or no differences in accuracy performance (when compared with the existing non-parallel DNN models). 
Future work will be focused on further improvement of parallelization approach for efficient training performance.

\section*{Acknowledgment}
The work reported in this paper has formed part of the project by Royal Society - Academy of Medical Sciences Newton Advanced Fellowship (NAF \textbackslash R1\textbackslash 180371).

\section*{References}


\end{document}